\PassOptionsToPackage{sort}{natbib}
\PassOptionsToPackage{inline}{enumitem}

\documentclass[paper]{sigirforum}

\def\pubissue{Vol. 59 No. 1 -- June 2025}

\usepackage{xcolor}
\usepackage{enumitem}

\begin{document}

\title{Information Retrieval for Climate Impact\\[1.1ex] \Large Report on the MANILA24 Workshop}

\authors{
\author[m.derijke@uva.nl]{Maarten de Rijke}{University of Amsterdam}{The Netherlands}
\and
\author[bart.vandenhurk@deltares.nl]{Bart van den Hurk}{Deltares}{The Netherlands}
\and
\author[flora.salim@unsw.edu.au]{Flora Salim}{UNSW Sydney}{Australia}
\and
\authorothers{Alaa Al Khourdajie,  
Nan Bai, 
Renato Calzone,
Declan Curran, 
Getnet Demil, 
Lesley Frew,
Noah Gie{\ss}ing, 
Mukesh Kumar Gupta, 
Maria Heuss, 
Sanaa Hobeichi, 
David Huard, 
Jingwei Kang, 
Ana Lucic,
Tanwi Mallick,
Shruti Nath, 
Andrew Okem,
Barbara Pernici, 
Thilina Rajapakse, 
Hira Saleem, 
Harry Scells, 
Nicole Schneider, 
Damiano Spina, 
Yuanyuan Tian, 
Edmund Totin,
Andrew Trotman, 
Ramamurthy Valavandan,
Dereje Workneh, 
Yangxinyu Xie}
}

\maketitle 

\begin{abstract}
The purpose of the MANILA24 Workshop on information retrieval for climate impact was to bring together researchers from academia, industry, governments, and NGOs to identify and discuss core research problems in information retrieval to assess climate change impacts. 
The workshop aimed to foster collaboration by bringing communities together that have so far not been very well connected -- information retrieval, natural language processing, systematic reviews, impact assessments, and climate science. The workshop brought together a diverse set of researchers and practitioners interested in contributing to the development of a technical research agenda for information retrieval to assess climate change impacts.
\end{abstract}

\section{Introduction}
Human-induced climate change, including more frequent and intense extreme events, has caused widespread adverse impacts and related losses and damages to nature and people, beyond natural climate variability~\citep{ipcc-2022-climate}. 
The Intergovernmental Panel on Climate Change (IPCC) is the leading international body for assessment of climate change. 
Approximately every six years, the IPCC releases an assessment report on the different aspects, drivers, and impacts of climate change based on an assessment of the literature. 
The IPCC report has contributions from three  different working groups. 
In particular, Working Group II (WGII) of the IPCC ``assesses the impacts, adaptation, and vulnerabilities related to climate change, from a world-wide to a regional view of ecosystems and biodiversity, and of humans and their diverse societies, cultures and settlements. 
It considers their vulnerabilities and the capacities and limits of these natural and human systems to adapt to climate change and thereby reduce climate-associated risks together with options for creating a sustainable future for all through an equitable and integrated approach to mitigation and adaptation efforts at all scales''~\citep{ipcc-2024-wgii}.

Fully using the available knowledge on emerging climate change impacts is key to informing global policy processes as well as regional and local risk assessments and on-the-ground action on climate adaptation~\citep{schleusssner-2020-scenarios}. 
The exponential growth in peer-reviewed scientific publications on climate change is pushing manual expert assessments to their limits~\citep{callaghan-2021-machine-learning-based,callaghan-2020-topography,joe-2024-assessing}. 
While literature aggregated on the level of continents or world regions might be useful to the global policy process, informing concrete climate adaptation typically requires localized and contextualized information on climate impacts~\citep{conway-2019-need}.
Tracking the effectiveness and progress of adaptation actions has proven difficult~\citep{sietsma-2024-next} -- any attempt to track adaptation progress will need to be capable of rapidly handling large and varied datasets and literature sources, while acknowledging highly localized and contextualized information.

This workshop report brings together researchers from the information retrieval, natural language processing, systematic review, and climate science communities in an attempt to develop an agenda to advance information retrieval for climate impact assessment. 
We begin by dissecting the problem: what is the information need addressed by the IPCC WGII (Section~\ref{section:the-need})? We then switch to methodologies, and in particular systematic reviews (Section~\ref{section:methodology}). We review different resources available (and/or needed) to support information retrieval for climate impact, including test sets and implementations (Section~\ref{section:resources}). We then address the question of how to make new technological advances in information retrieval work as part of the IPCC WGII assessment workflow (Section~\ref{section:usage}). The paper concludes with a broader perspective. To remain focused, our discussion and analysis is centered around IPCC WGII and its mission -- we believe, however, that many of our questions and suggestions have the potential to contribute to the workflows of other IPCC working groups and task forces.

\section{The Information Need}
\label{section:the-need}
The IPCC aims to provide governments at all levels with scientific information that they can use to develop climate policies~\citep{ipcc-2024-about}. The IPCC is divided into three working groups (WGs). WGI deals with the physical science basis of climate change, WGII with climate change impacts, adaptation and vulnerability, and WGIII with mitigation of climate change.
The IPCC does not conduct its own research, run models, or make measurements of climate or weather phenomena. Its role is to assess the scientific, technical and socio-economic literature relevant to understanding climate change, its impacts, future risks, and options for adaptation and mitigation. 
Author teams assess all such information from any source that is to be included in the report.

As pointed out in the introduction, approximately every six years, the IPCC releases a series of reports on the different aspects of climate change based on large-scale assessment of all the latest literature. 
In early 2022, as part of the IPCC’s sixth assessment report (AR6), the IPCC released the report of WGII on impacts, adaptation and vulnerability, which ``assesses the impacts of climate change, looking at ecosystems, biodiversity and human communities at global and regional levels. It also reviews vulnerabilities and the capacities and limits of the natural world and human societies to adapt to climate change'' \citep{ipcc-2022-climate}. The report covers ecosystems, sectors, sustainable development goals, and regions, an integrated technical summary, and a summary for policymakers. The AR6 WGII report pulls together evidence and findings from more than 34,000 journal papers and reports; it is written by 270 authors from 67 countries.

The assessment report currently in development (AR7) faces a twin challenge: assessing and synthesizing a fast-growing amount of literature related to climate change, and addressing relevant topics and areas where little data is available in the published literature. 
We believe that information retrieval can potentially help address both aspects, by (i) helping IPCC author teams sift through the literature, and (ii) exposing relevant information previously inaccessible to author teams due to various barriers (language, etc.).

\subsection{Proposed Research}
\label{subsection:the-need-proposed-research}

We propose an agenda driven by the ambition to produce open evidence synthesis, based on principles of transparent information gathering, curation, traceability, and information quality.  Fairness and mitigating different types of bias (e.g., geographic, cultural) are important proposed research lines. 
Specific attention is to be paid to underrepresented sources of information and communities.
Existing AI tools work well with large amounts of data, so have the potential to exacerbate imbalances in data and literature coverage~\citep{bahri-2024-explaining}.

How do we make sure AI can help with topics or areas where less information is available? Another line of research has to do with the data sources used for evidence synthesis: 
(i)~(capturing and) using domain-specific information, 
(ii)~the usage of grey literature, i.e., the diverse and heterogeneous body of material available outside, and not subject to, traditional academic peer-review processes such as preprints and policy documents~\citep{adams-2017-shades},
(iii)~Indigenous Knowledge and local knowledge, especially if it is not available online, and 
(iv)~(qualitative) data collected through grass roots and/or citizen data science efforts.

It is important that retrieval and analysis models can handle multi-modal data (i.e., text, numerical, images, etc.). Only published data currently goes into the IPCC review process, and this is will not cover, e.g., remote sensing data that has no description and interpretation in the literature. The final direction concerns important guardrails that apply to the evidence synthesis process. One is that there is zero tolerance for black boxes -- this is important to avoid undermining the credibility of the synthesis with policy makers and governments. 
An important requirement for ``unconventional'' sources of information is its quality assessment; IPCC authors need to defend the appropriateness of the use of non-peer reviewed literature; we need to elaborate on this quality assessment of non peer-reviewed literature.
And another concerns the carbon emissions when choosing information retrieval (IR) approaches (i.e., energy requirements of using large language models (LLMs)). Do benefits of the model/data/research outweigh the carbon costs?

\subsection{Research Challenges}
\label{subsection:the-need-challenges}

These general research questions manifest themselves along the entire evidence synthesis chain and motivate a broad range of concrete research directions to be investigated concerning the information need that IPCC addresses. Some of these have to do with the scope of the evidence synthesis, some concern data acquisition and quality assessment, and some with contextual factors.

\paragraph{Data gathering and sharing.} If more diverse sources of evidence are integrated into the evidence synthesis process, how can we ensure open, transparent data collection and sustainable sharing?

\paragraph{Data quality.} A common suggestion is that evidence should include more diverse sources. 
Evidence based on solely grey literature should be supported by other evidence lines. However, combining datasets of different structure or manually annotating unstructured data is time-intensive, and grey literature in particular is difficult to work with~\citep{sietsma-2024-next}. How can local expert knowledge be embedded in the evidence synthesis process, while ensuring that data/literature imbalances are not exacerbated? How can acceptable data be created from unconventional sources?

\paragraph{Climate adaptation.} It has been claimed that robust synthesis of climate \emph{adaptation} literature and insights was under-represented in climate impact assessments, including the most recent version of the WGII report~\citep{berrang-ford-2021-systematic}. How to address and capture successful adaptations, whether technological, institutional, behavioral, or nature-based~\citep{oneill-2022-key}. 
How to surface and flag information that helps avoid ``maladaptations'' that may increase vulnerability, lock-in, or unequal impacts~\citep{barnett-2013-minimising}? 
How to measure and mine outcomes of adaptation actions and aggregate them, e.g., based on similarity of contexts, precision, causality?

\paragraph{Timeliness.} A single IPCC assessment cycle lasts approximately six years and the resulting syntheses are difficult to keep up to date. How can more recent statistics, figures, and findings from research papers lead to ``updated'' reports? What would acceptable ``living evidence'' in the context of the IPCC look like~\citep{elliott-2017-living}?

\paragraph{Evaluation.} How can we evaluate ``fit to information need'' of assessment reports or of the various summaries, such as the summary for policymakers~\citep{ipcc-2022-summary} or the technical summary~\citep{portner-2022-technical}, that are generated based on the reports? The IPCC uncertainty guidances notes~\citep{mastrandrea-2010-guidance} are meant to assist in the consistent treatment of uncertainties in developing expert judgments and in communication -- how can they be integrated with traditional evaluation guidelines from the IR community?

\subsection{Obstacles and Risks}
\label{subsection:the-need-obstacles-and-risks}

To enable this research we need broad collaborations between IR researchers and climate impact researchers. Finding effective ways of working together and finding a shared vocabulary requires considerable effort that some researchers may not be able to afford or have insufficient institutional support, resources or recognition for~\citep{allan-report-2018}. 
An important risk of using information retrieval based literature collection concerns the continuation or even exacerbation of imbalances in coverage in IPCC reports by focusing on areas and topics that have large amounts of data available for training. A potential obstacle for bringing recent advances in IR technology to bear on the IPCC's evidence synthesis workflow is the broad -- and successful -- uptake of technologies that are often referred to as ``opaque'' or ``black-box'' technologies such as LLMs and generative AI. Another key constraint is that the IPCC report production process is guided by a set of principles agreed by the panel,\footnote{\url{https://www.ipcc.ch/site/assets/uploads/2018/09/ipcc-principles-appendix-a-final.pdf}} potentially limiting the integration of important innovations in the evidence synthesis process.

\section{Methodology}
\label{section:methodology}

The logic of the IPCC assessments is somewhat similar to the workflow of systematic reviews. 
Traditional systematic reviews are structure following a number of, more or less, standardized steps~\citep{cooper-2018-defining}: 
(i) querying (which includes
(a) protocol definition and
(b) search strategy development),
(ii) screening (which includes
(a) study abstract screening and
(b) study full-text screening),
and
(iii) aggregating (which includes
(a) study synthesis and results preparation and
(b) dissemination of systematic review).
IPCC's approach can be characterized using the following dimensions:

\begin{description}
	\item[Rigorous assessment:] Use established methods for comprehensive reviews, aligning with IPCC’s mission to assess climate literature.
	\item[Database utilization:] Employ robust academic databases such as Scopus and Web of Science for extensive literature search.
	\item[Literature screening:] Implement a systematic selection process to ensure the inclusion of relevant studies.
	\item[Publication coding:] Use qualitative analysis software to code and categorize information effectively.
	\item[Scholarly perspectives] Evaluate the gathered facts and identify research gaps to provide a well-rounded scholarly insight.
	\item[Data visualization:] Create visual representations such as diagrams, word clouds, and tables to present the synthesized data clearly.
	\item[Connect with practitioners:] Summarize key findings to inform policy implications and engage with practitioners for real-world interpretation.
\end{description}

\noindent%
IPCC's current approach to evidence synthesis has several shortcomings:
\begin{enumerate*}[label=(\roman*)]
\item the growth of the literature: human-driven information collection skews search relevance, hence tool support is needed to manage and streamline new literature; 
\item information comes in diverse formats, both structured and unstructured: fragmented information hinders understanding and analysis, hence tool support is needed to integrate structured and unstructured data; and
\item there is a significant time lag due to the six year cycle, for which \emph{living} systematic reviews may be a way forward~\citep{elliott-2017-living}.
\end{enumerate*}

\subsection{Proposed Research}
\label{subsection:methodology-proposed-research}

Developing evidence synthesis methodologies that help address challenges in IPCC's assessment process requires new research on a broad range of information retrieval topics subject to a number of constraints and guardrails. In particular, information retrieval methods are needed for identifying, integrating, understanding, and selecting relevant literature and information extraction methods for (i) efficiently synthesizing relevant data and (ii) improving accuracy and relevance of extracted information. 
The main tasks requirements that these methods need to address are 
\begin{enumerate*}[label=(\roman*)]
\item eliminating hallucination in case LLM-based methods are used;
\item mitigating bias by enhancing the selection process and promoting fairness,  inclusivity, and representativity; and
\item maintaining human accountability and ensure transparency in AI-assisted processes while securing a comprehensive coverage of the literature to be assessed.
\end{enumerate*}

\subsection{Research Challenges}
\label{subsection:methodology-challenges}

Challenges are faced on each of the areas that the proposed research covers. The proposed research touches on the core of evidence synthesis: the way of working, the data, data enrichment, the synthesis and summarization process,  its timeliness, and the role of people in the process.

\paragraph*{Protocol development} How should evidence synthesis protocols be co-designed to facilitate  collaborative research, adhering to Indigenous protocols, and including under-represented, culturally and linguistically diverse groups~\citep{roberts-2021-guide}? What are effective ways of engaging with these communities? 
How can we use generative AI to produce effective communication material for different communities~\citep{lansbury-2023-aboriginal}?

\paragraph*{Characterizing the literature} If the input to the synthesis process diversifies, what is the process of search? How, using which credibility criteria, should we determine corpora of scientific and ``non-traditional'' research outputs (e.g., Indigenous Knowledge), grey literature (e.g., government reports), and select sources to include? 

\paragraph*{The labeling bottleneck} Traditionally, assessment reports relied on domain experts who identified and selected, often in teams, documents for relevance and for specific aspects of climate impact as part of the evidence gathering and synthesis process. The rapid growth of scientific literature has made this increasingly challenging. Can evidence gathering and screening be automated with recent advances in the use of LLMs for relevance labeling \citep{thomas-2024-large}? Can information extraction tasks be off-loaded to LLMs~\citep{mallick-2024-analyzing}? And to which degree can LLMs be used to compare and synthesize extracted evidence~\citep{joe-2024-assessing}?

\paragraph*{Trustworthy synthesis} IPCC's assessment reports inform policy-makers. They provide a scientific basis for governments at all levels to develop climate related policies.
The importance of understanding how to synthesize information with varying degrees of fidelity cannot be overstated. 
For example, in the climate model intercomparison project\footnote{CMIP, \url{https://wcrp-cmip.org/cmip-phases/cmip6/}}  organizations can submit their climate models, which others can use to do climate simulations, usually by compute averages over all the different models. In CMIP6, there were 50+ models; some were ``too hot'' \citep{hausfather-2022-climate} and averaging would therefore overestimate temperature increase. IPCC6 acknowledged this, stating that models should be weighted according to fidelity, not just blindly synthesized.

IPCC's reports, and the processes and systems underlying them, need to be trustworthy~\citep{wang-2024-zero-shot}. From a technological point of view, a process or technology can be trustworthy for \emph{intrinsic} reasons, e.g., because a sufficient level of transparency of the process and technology are provided. Or it can be trustworthy for \emph{extrinsic} reasons, i.e., because we have ways of probing it for properties such as accuracy, reliability, repeatability, resilience, and safety. How can these intrinsic and extrinsic approaches be developed for large-scale evidence synthesis?

\paragraph*{Living evidence} Evidence changes over time. What are effective update protocols? The temporal dimension plays a big part in evidence outcomes. Is the amount of literature that is to be assessed in a given IPCC cycle still manageable and can it be expected to be representative for the target time period? What would the requirements be for self-updating synthesis systems? How should we preserve previous versions~\citep{simmonds-2017-living}?

\paragraph*{Human in the loop synthesis} How can we measure the cost, and benefit, of having humans in the evidence synthesis loop~\citep{thomas-2017-living}? How should we provide feedback to the systems and who provides this feedback? And vice versa, how do we transfer tasks from systems to people, or trigger interventions from people in complex evidence synthesis tasks? Can interactive or collaborative frameworks from other tasks be adopted~\cite[e.g.,][]{zhang-2021-human-machine}?

\subsection{Obstacles and Risks}
\label{subsection:methodology-obstacles-and-risks}

We  currently know very little about the extent to which automation of key evidence synthesis steps such as query automation, screening automation, and automation of the aggregation results are effective for evidence synthesis in the climate impact domain, and what little we know has usually been learned on English language scientific publications. An important obstacle is the lack of data and benchmarks (especially Indigenous Knowledge, and low-resource languages) and lack of training corpus for these languages. An important risk is limited success in community engagement, across disciplines and across different types of stakeholders. It is important to do this in the right way by adhering to responsible practices, even though they may not have been established for all types of (non-traditional) domain experts and underrepresented groups~\citep[see, e.g.,][]{lewis-2020-indigenous}.

\section{Resources}
\label{section:resources}

Progress in the development of effective methods to automate all steps in systematic reviews has been greatly facilitated by the availability of shared resources. In evidence synthesis in healthcare and medicine, for instance, we have witnessed an explosion of methods due to shared tasks and datasets. Prominent examples are the CLEF Technology Assisted Reviews task~\citep{kanoulas-2019-clef} and the CSMeD: Meta-dataset for systematic review automation evaluation~\citep{kusa-2023-csmed}. In information retrieval for climate impact there is an urgent need for similar standardized datasets and test collections.

\subsection{Proposed Research}
\label{subsection:resources-proposed-research}

Support for the creation of shared resources (data, code, evaluation methods) requires an understanding of the information and steps that are needed to accomplish the assessment goals, of the (inter)actions of evidence reviewers working towards the goals.
This has two sides:
(i) collecting information on current assessment practices, and
(ii)~analyzing data, process and options for automation.

It is instructive to revisit the main steps in the systematic review process~\citep{scells-2021-query,cooper-2018-defining} -- \emph{querying}, \emph{screening}, and \emph{aggregating} -- and see what progress has been enabled using shared resources and large language models in recent years.

The \emph{querying stage} typically involves steps such as query formulation, query refinement, and query representation. As an example of recent progress using LLMs and building shared resources in the context of querying, \citet{wang-2023-can} explore the use of LLMs in the medical domain to follow instructions and generate queries with high-precision, while trading this off for recall. Using resources from the CLEF initiative\footnote{\url{https://www.clef-initiative.eu}} for evaluation, the authors find that LLMs are a valuable tool for rapid reviews where time is a constraint and trading off higher precision for lower recall is acceptable.

The \emph{screening} stage typically involves screening prioritization, cut-off prediction, and document classification. As another example of recent progress facilitated by the CLEF e-Health,\footnote{\url{ https://github.com/CLEF-TAR}} the Text Retrieval Conference (TREC) Total Recall benchmarking activity,\footnote{\url{http://plg.uwaterloo.ca/~gvcormac/total-recall/}} and the TREC Legal\footnote{\url{https://trec-legal.umiacs.umd.edu/}} resources, \citet{stevenson-2023-stopping} propose a stopping method based on point processes and test the robustness of a wide range of statistical and neural stopping methods in a series of contrastive experiments.

Finally, the \emph{aggregating} stage typically involves information extraction, result synthesis, statistical analyses, and dissemination. As an example of recent progress that uses LLMs and builds on shared resources, \citet{shaib-2023-summarizing} examine the potential of LLMs to synthesize (that is, aggregate the findings presented in) multiple medical articles in a way that accurately reflects the evidence. Using articles indexed in the Trialstreamer database\footnote{\url{https://trialstreamer.ieai.robotreviewer.net}} and meta-analyses from the Cochrane Library,\footnote{\url{https://www.cochranelibrary.com/}} the authors conclude that human-written summaries in the (bio)medical domain require a level of synthesis that is not yet captured by today's LLMs.

Do these findings generalize from the medical domain to information retrieval for climate impact and to all aspects of the systematic review workflow? What are the resources needed to facilitate similar advances in information retrieval for climate impact and how do we create them?

\subsection{Research Challenges}
\label{subsection:resources-challenges}

Resources in support of information retrieval for climate impact can be addressed at  two levels -- at the level of individual resources and at the aggregate level of multiple resources.
First level, at the level of individual resources, we see the following research challenges.

\subsubsection{Individual resources}

\paragraph*{Document collections}
Several collections relevant to support research into information for climate impact have been used in recent years. White literature collections used include  a subset of the Semantic Scholar Open Research Corpus consisting of 600K climate-related articles~\citep{mallick-2024-analyzing}, subsets (abstracts, titles and
metadata (no full text)) from the Web of Science Core Collections databases consisting of 375K~\citep{callaghan-2020-topography} and 565K~\citep{sietsma-2021-progress} articles. 
OpenAlex is another option.\footnote{\url{https://openalex.org/}.}
Grey literature of different sorts has also been used in the context of climate science, even though for IPCC not all types of grey literature (e.g., tweets and media coverage) are considered eligible. 
\citet{bai-2024-inferring} share a collection of 60K (multi-modal) tweets covering different climate change stances. 
The GDELT climate news narrative datasets, covering 2009 to 2020, represent television, language, the global perspective and context dating back from half a decade to a decade in all.\footnote{\url{https://blog.gdeltproject.org/four-massive-datasets-charting-the-global-climate-change-news-narrative-2009-2020/}} 
\citet{frew-2024-retrogressive} use web archives and query logs to demonstrate how conflicting climate change stances between government administrations were correlated with content drift and lack of public access to climate information webpages over time.
How can we create a large-scale, heterogeneous corpus for climate impact, a corpus, moreover that would support a heterogeneous benchmark containing diverse climate-related tasks (like the BEIR dataset in ad-hoc retrieval~\citep{thakur-2021-beir})? How do we determine the usefulness of including non-peer reviewed scientific reports, theses, policy documents in corpus and benchmark development? How can we create benchmarks, for both static and tracking tasks? 

\paragraph*{Knowledge sources}
Knowledge sources play an important role in making sense of both white and grey climate literature. Example knowledge sources used to understand climate science literature at scale include KnowWhereGraph~\citep{janowicz-2022-know}, a geo-knowledge graph that is based on existing standards like RDF,\footnote{\url{https://www.w3.org/RDF/}} OWL\footnote{\url{https://www.w3.org/OWL/}} and GeoSPARQL,\footnote{\url{https://www.ogc.org/publications/standard/geosparql/}} incorporates custom ontologies, and uses a hierarchical grid for spatial representations; it is built upon many open-source spatial datasets, including an expert knowledge graph built using scientific publication/papers on disaster relief.\footnote{\url{https://knowwheregraph.org/graph/}} Two other important knowledge sources for anomaly and unusual event detection are the iNaturalist biological species dataset\footnote{\url{https://www.inaturalist.org/observations?place_id=any&subview=map}} and the local environmental observer (LEO) network.\footnote{\url{https://www.leonetwork.org/en/docs/about/about}} How can we use these resources to enrich white and grey literature at scale and enable different types of semantic search~\citep{balog-2018-entity}?

\paragraph*{Tools and technology}
Various domain-specific resources and tools have been developed for information retrieval for climate impact recently. For example, 
ClimateBert is transformer-based language model that is further pretrained on over 2 million paragraphs of climate-related texts, crawled from various sources such as common news, research articles, and climate reporting of companies~\citep{webersinke-2021-climatebert}.
WildfireGPT is an LLM designed to support the analysis of wildfire data, by a diverse set of end users, including  researchers and engineers. WildfireGPT is an interesting example of a domain-specific resource not just because of its use cases but also because it uses a broad range of domain-specific data~\citep{xie-2024-wildfiregpt}.
\citet{thulke-2024-climategpt} introduce ClimateGPT, a family of large language models  designed for understanding and responding to information about climate change; the paper demonstrates the effectiveness of adapting a strong general-purpose LLM through continued pre-training on a curated climate corpus and fine-tuning it with high-quality, human-expert-involved instruction data.

\citet{tian-2025-advancing} propose a retrieval-reranking framework that uses LLMs to identify and recommend semantically and spatiotemporally similar unusual environmental events described in news articles and web posts. By integrating embedding-based semantic search with a geo-time re-ranking strategy that integrates multi-faceted criteria including spatial proximity, temporal association, semantic similarity, and category-instructed similarity, the work demonstrates how LLMs can support climate event mining in a real-world platform setting.
\citet{mallick-2024-analyzing} describe an end-to-end decision-support tool that accurately identifies location information within textual documents, enabling extraction of geographic-specific climate trends and topics, to provide a concise understanding of geographic-specific climate change trends. How can tools and technology with such far-reaching potential be made accessible to everyone, not just the technologically privileged?
\citet{rajapakse-2024-simple} describe a widely used library designed to simplify the training, evaluation, and usage of transformer models that operationalizes the idea of ``open-source for all'' with adoption in materials science, environmental science, and healthcare. Can we put this perspective to work for all tools and technology (or technology components) that help make sense of information related to climate impact?

\subsubsection{Aggregating across multiple resources}

There are multiple research challenges that concern the development or usage at different levels of aggregation.

\paragraph*{Transfer \& generalization}
The first challenge concerns the use of the broad range of advances and resources realized in language technology in recent years. How can we transfer these advances to the domain and tasks of information retrieval for climate impact with minimal effort to reduce costs and accelerate the uptake \emph{while} achieving reliable performance?
Can we take the findings from evidence synthesis in technology assisted reviews in the medical or legal domain and transfer these to the climate domain? Can these questions be turned into few-shot learning problems~\citep{wang-2025-cooperative}? 

\paragraph*{Benchmarking}
The lack of standardized benchmarks and datasets for validating and comparing information retrieval models for climate impact poses significant challenges to the development, assessment, and comparison of information retrieval technologies in this field. Can we use (very) limited human labeling and rely on LLM-generated labels where large quantities of labels are needed~\citep{thomas-2024-large} for test collection creation? To which degree would a weak-to-strong alignment perspective~\citep{lyu-2025-macpo} be helpful for creating the required benchmarking resources?

\paragraph*{Stacking up}
Recent years have witnessed great progress in the performance of individual components that make up a systematic review workflow -- in querying, screening, and aggregating. 
Do improvements in the components lead to improvements in the overall assessment process? How do issues such as bias, reproducibility, and performance emerge in complex retrieval for impact pipelines~\citep{hocking-2023-overcoming}? How can the full review pipeline be optimized effectively? Is the field ready to explore end-to-end learnable approaches to information retrieval for climate impact such as generative information retrieval~\citep{tang-2024-recent-sigir}?

\paragraph*{Scaling up}
Managing and processing the vast amounts of heterogeneous data from diverse sources such as sensors, climate models, literature and satellite data poses significant challenges. In the medical domain, organizations such as Cochrane and the Joanna Briggs Institute offer extensive methodological guidance for dealing with large-scale collections, putting a large emphasis on the use of software such as Covidence\footnote{\url{https://www.covidence.org}} or JBI Sumari.\footnote{\url{https://sumari.jbi.global}} How do such tools scale up to the size and heterogeneity of materials related to climate impact?
Furthermore, linking text data from white and grey literature that describes a climate event to relevant physical measurements about the event, like temperature, rainfall, audio-visual recordings, etc. at a world-wide scale is a challenge that goes far beyond the entity linking challenges that the information retrieval community addresses today~\citep{meij-2013-entity}.

\subsection{Obstacles and Risks}
\label{subsection:resources-obstacles-and-risks}

Collaboration among researchers in information retrieval, natural language processing, data engineering, and climate science, developers, and methodologists is essential for advancing the field of information retrieval for climate impact. The fact that these communities and sub-communities -- with different sharing and evaluation cultures -- are largely disconnected, is a risk that may hinder progress in the development and uptake of resources and benchmarks.

\section{Usage}
\label{section:usage}
Recall the core steps from IPCC's overall evidence synthesis process -- querying, screening, and aggregating. In each of these steps it employs provenance tracking as a core component of the workflow. By documenting the origin and reliability of the evidence, WGII provides a foundation for informed decision-making on climate change impacts, adaptation, and vulnerability. 

\subsection{Proposed Research}
\label{subsection:usage-proposed-research}

IPCC's way-of-working (with a strong emphasis on provenance tracking) comes with several key challenges. First of all, the exponential growth and increasing complexity of scientific literature complicates the IPCC's mandate to conduct ``comprehensive, objective, open and transparent'' assessment in line with the IPCC principles, an issue that is exacerbated by the workload required from scientists to engage with assessment~\citep{de-gol-2023-broadening,minx-2017-learning}. Second, the length of the IPCC reports poses significant challenges to access the detailed underlying evidence in the main reports. For example, in the most recent assessment round (AR6), the two main reports exceed 2,000 pages each, with one exceeding 3,000 pages. Including additional specialized reports and summaries, a total of ${\sim}$12,000 pages was reached~\citep{al-akhourdajie-2024-role}.
The sheer volume hinders the effective dissemination of important scientific findings that do not feature in the summaries.

Assuming that the community manages to make progress along the lines suggested in Section~\ref{section:the-need} (on understanding IPCC's information need), Section~\ref{section:methodology} (on methods for addressing IPCC's information needs), and Section~\ref{section:resources} (on resources to implement and assess those methods), how and where can  the resulting advances be made to work for IPCC's assessments, given the challenges outlined above?
We need to navigate the vast and ever-expanding corpus of white and grey literature on climate in an efficient manner to enable comprehensive and credible consensus building processes.
How to use tools to effectively communicate the extensive findings in the lengthy IPCC reports while maintaining the integrity of the assessments?
How to determine the weight of heterogeneous evidence, complexity, uncertainty and the level of confidence?

\subsection{Research Challenges}
\label{subsection:usage-challenges}

\citet{al-akhourdajie-2024-role} summarize several current lines of work to structure the assessment work, in a way that addresses the questions listed above and that provides connecting points for the agendas in Section~\ref{section:the-need}, \ref{section:methodology} and~\ref{section:resources}. The IPCC scoping process could use \emph{evidence maps} to highlight key topics. Once outlines are agreed, \emph{topic discovery} could help authors of assessment reports plan drafts by clustering publications and topics. A \emph{hybrid exploration} approach could then be used to inform assessment drafting. \emph{Collaborative platforms} could then be used to expand engagement, while \emph{exploration} tools would help to expand coverage.

\paragraph{Synthesis and evidence maps}
An evidence map is a visual representation of the available research on a particular topic.
It provides an overview of the extent, range, and nature of evidence, highlighting where research exists and, importantly, where gaps remain~\citep{hempel-2014-evidence}. 
In the area of climate adaptation research gaps are present in the assessment of adaptation effectivity; there is regional skewness in literature coverage (with substantially more literature in global north countries), and there is poor coverage of quantitative assessments of adaptation performance~\citep{ipcc-2022-climate}.
How can LLMs be used to identify and classify white and grey literature publications and provide rapid and comprehensive coverage?

\paragraph*{Topic and event discovery}
How can traditional information retrieval and LLM-based methods be used to identify and track topics and events in white and grey literature so as to reveal gaps?

\paragraph*{Hybrid exploration}
Expert judgment remains crucial for assessing literature that is meant to help inform policy-makers.
Solely relying on human expert judgment introduces big risks in skewed assessments, as the volume of literature gets simply too large to be reviewed systematically. Hence, it makes sense to move forward with hybrid approaches, in order to meet the IPCC goals of comprehensive and unbiased assessments. 
How can we organize effective iterative human-in-the-loop (or machine-in-the-loop) approaches that alternate between query reformulation and retrieval steps and are likely to identify important thematic clusters and yield content-based insights?

\paragraph*{Collaborative
platforms}
\citet{de-gol-2023-broadening} describe their experience with a collaborative technology platform, tailored to support an updated process of elaborating IPCC reports.
Among other things, the platform, ScienceBrief, featured a living map of evidence that used motion and spatial reasoning through animation and physical simulation to visualize the scientific consensus around a topic.
\citeauthor{de-gol-2023-broadening}\ argue that such platforms could greatly enhance IPCC assessments by making them more open and accessible, further increasing
transparency. What sort of content-based recommendation methods, based on ideas in Sections~\ref{section:methodology} and~\ref{section:resources}, could help to create engagement?

\paragraph*{Exploration}
How can modern information retrieval methods, in combination with visualization and semantic tools, support the discovery of publication networks, enable swift navigation and simplify subsequent literature synthesis?

\subsection{Obstacles and Risks}
\label{subsection:usage-obstacles-and-risks}

Beyond the data, technological, and methodological challenges listed above, the uptake of outcomes from the research suggested in Section~\ref{section:the-need}, \ref{section:methodology}, and \ref{section:resources} in climate impact assessments faces ethical and social challenges~\citep{ghosh-2022-co-development}.
Ensuring that the increased use of advanced information retrieval technology promotes equity and justice is essential -- often, the people most affected by climate change are the least involved in the creation of the technologies, or the assessments of climate change.
Avoiding the creation of new inequalities, e.g., in terms of access to information, contribution to the assessment, usefulness of the IPCC products, or exacerbating existing ones, is a key consideration.
Ensuring transparency and accountability in the use of new information retrieval technologies is crucial for building trust and legitimacy.
This includes clearly documenting the methods, data, and assumptions used in the assessment.
Finally, engaging stakeholders in the development and use of new information retrieval technologies is essential for ensuring that assessments are relevant and useful.
This requires effective communication and collaboration.

\section{Final Note}
\label{section:perspective}
Climate change is real. 
The information retrieval community has a unique opportunity to inform and foster a collaborative ecosystem of researchers and practitioners that contribute to effective information retrieval solutions in support of research and decision-making to help address the reality of climate change impacts~\citep{sietsma-2024-next}.
The agenda setting activities of the MANILA24 workshop were meant to do just that. Preparations for MANILA25, the second edition of the workshop, to be held at SIGIR 2025, are well under way, continuing the goal of developing, maintaining, and executing an effective research agenda for information retrieval for climate change impacts.

\section*{Acknowledgements}
We would like to thank the organizers of SIGIR 2024 for hosting the MANILA24 workshop.

AAK was supported by the European Union's Horizon Europe research and innovation programme (101056306, IAM COMPACT).
GD was funded by the Academy of Finland Digital Water Flagship Project, European Chist Era Waterline Project.
LF was funded by the Old Dominion University Division of Student Engagement \& Enrollment Services, Graduate Student Travel Award.
MH was funded by the Dutch Research Council (024.004.022).
SH acknowledges the support of the Australian Research Council Centre of Excellence for the Weather of the 21st Century (CE230100012).
JK was supported by the Dutch Research Council (KICH3.LTP.20.006).
TM acknowledges support by Laboratory Directed Research and Development (LDRD) funding from Argonne National Laboratory, provided by the Director, Office of Science, of the U.S.\ Department of Energy (DEAC02-06CH11357).
TR acknowledges the support of the Dreams Lab, a collaboration between Huawei Finland, the University of Amsterdam, and the Vrije Universiteit Amsterdam.
MdR was funded by the Dutch Research Council (024.004.022, NWA.1389.20.183, KICH3.LTP.20.006) and the European Union's Horizon Europe research and innovation program (101070212, FINDHR).
HS is supported by SmartSat CRC.
FS acknowledges the support of the ARC Centre of Excellence for Automated Decision-Making and Society (ADM+S, CE20010000) and SmartSat CRC.
DS was funded by the Australian Research Council (DE200100064) and the ARC Centre of Excellence for Automated Decision-Making and Society (ADM+S, CE200100005).
YT was funded by the U.S. National Science Foundation (2230034, 2120943, and travel award for SIGIR), Google.org’s Impact Challenge for Climate Innovation Program, and OpenAI’s Researcher Access Program.

All content represents the opinion of the authors, which is not necessarily shared or endorsed by their respective employers and/or sponsors.

\appendix

\appendixauthorothers
\label{appendix:authors}
This is a list of authors of this publication. 
Workshop organizers:
\begin{itemize}
	\item Maarten de Rijke (University of Amsterdam, The Netherlands)
    \item Bart van den Hurk (Deltares, The Netherlands)
	\item Flora Salim (UNSW Sydney, Australia)
\end{itemize}
Others who contributed to this publication as well:
\begin{itemize}
	\item Alaa Al Khourdajie (Imperial College London, UK and International Institute for Applied System Analysis (IIASA), Austria; workshop participant)
	\item Nan Bai (Delft University of Technology and Wageningen University and Research, The Netherlands; workshop participant)
	\item Renato Calzone (ILUSTRE/Tilburg University, The Netherlands; program committee member)
	\item Declan Curran (UNSW Sydney, Australia; workshop participant)
	\item Getnet Demil (University of Oulu, Finland; workshop participant)
	\item Lesley Frew (Old Dominion University, USA; workshop participant)
	\item Noah Gie{\ss}ing (FIZ Karlsruhe/MARDI, Germany; workshop participant)
	\item Mukesh Kumar Gupta (Singapore Management University, Singapore; workshop participant)
	\item Maria Heuss (University of Amsterdam, The Netherlands; program committee member)
	\item Sanaa Hobeichi (UNSW Sydney, Australia; program committee member)
	\item David Huard (Ouranos, Canada; workshop participant)
	\item Jingwei Kang (University of Amsterdam, The Netherlands; workshop participant)
	\item Ana Lucic (University of Amsterdam, The Netherlands; program committee member)
	\item Tanwi Mallick (Argonne National Laboratory, USA; workshop participant)
	\item Shruti Nath (AOPP/University of Oxford, UK; program committee member)
	\item Andrew Okem (Deltares, The Netherlands; workshop participant)
	\item Barbara Pernici (Politecnico di Milano, Italy; workshop participant)
	\item Thilina Rajapakse (University of Amsterdam, The Netherlands; workshop participant)
	\item Hira Saleem (UNSW Sydney, Australia; workshop participant)
	\item Harrisen Scells (University of T{\"{u}}bingen, Germany; workshop participant)
	\item Nicole Schneider (University of Maryland, USA; workshop participant)
	\item Damiano Spina (RMIT University, Australia; workshop participant)
	\item Yuanyuan Tian (Arizona State University, USA; workshop participant)
	\item Edmund Totin (Universit\'{e} Nationale d’Agriculture and World Vegetable Center; workshop participant)
	\item Andrew Trotman (University of Otago, New Zealand; workshop participant)
	\item Ramamurthy Valavandan (Deltares, The Netherlands; workshop participant)
	\item Dereje Workneh (Howard University, USA; workshop participant)
	\item Yangxinyu Xie (University of Pennsylvania, USA; workshop participant)
\end{itemize}

\section{Process}
This paper grew out of the half-day SIGIR 2024 workshop on Information Retrieval for Climate Impact~\citep{vandenhurk-2024-manila24}, which was held at SIGIR 2024 on July 18, 2024. 
The workshop was organized by BvdH, MdR, FS, who invited a number of researchers from the information retrieval and climate science domains to form the program committee. The workshop itself consisted of two parts. The first part was focused on gaining a better understanding of the information retrieval challenges faced by IPCC WGII, with four presentations on the topics underlying the four core sections of this paper (Sections~\ref{section:the-need}--\ref{section:usage}) as well as eight presentations that were selected based on an open call for contributions. The second part of the workshop consisted of a collaborative writing session, with four groups working on the topics underlying the four core sections of the paper, using ``seed topics'' contributed by members of the program committee. The workshop organizers used the seed topics, presentations, and outputs of the writing session to draft a first version of the paper, which was then shared with the author team for review and editing.

\bibliography{references}

\begin{thebibliography}{58}
\providecommand{\natexlab}[1]{#1}
\providecommand{\url}[1]{\texttt{#1}}
\expandafter\ifx\csname urlstyle\endcsname\relax
  \providecommand{\doi}[1]{doi: #1}\else
  \providecommand{\doi}{doi: \begingroup \urlstyle{rm}\Url}\fi

\bibitem[Adams et~al.(2017)Adams, Smart, and Huff]{adams-2017-shades}
Richard~J. Adams, Palie Smart, and Anne~Sigismund Huff.
\newblock Shades of grey: Guidelines for working with the grey literature in
  systematic reviews for management and organizational studies.
\newblock \emph{International Journal of Management Reviews}, 19\penalty0
  (4):\penalty0 432--454, 2017.

\bibitem[Al~Khourdajie(2024)]{al-akhourdajie-2024-role}
Alaa Al~Khourdajie.
\newblock The role of artificial intelligence tools in climate change
  scientific assessments.
\newblock \emph{SSRN preprint 4747126}, 2024.

\bibitem[Allan et~al.(2018)Allan, Arguello, Azzopardi, Bailey, Baldwin, Balog,
  Bast, Belkin, Berberich, von Billerbeck, Callan, Capra, Carman, Carterette,
  Clarke, Collins-Thompson, Craswell, Croft, Culpepper, Dalton, Demartini,
  Diaz, Dietz, Dumais, Eickhoff, Ferro, Fuhr, Geva, Hauff, Hawking, Joho,
  Jones, Kamps, Kando, Kelly, Kim, Kiseleva, Liu, Lu, Mizzaro, Moffat, Nie,
  Olteanu, Ounis, Radlinski, de~Rijke, Sanderson, Scholer, Sitbon, Smucker,
  Soboroff, Spina, Suel, Thom, Thomas, Trotman, Voorhees, de~Vries, Yilmaz, and
  Zuccon]{allan-report-2018}
James Allan, Jaime Arguello, Leif Azzopardi, Peter Bailey, Tim Baldwin,
  Krisztian Balog, Hannah Bast, Nick Belkin, Klaus Berberich, Bodo von
  Billerbeck, Jamie Callan, Rob Capra, Mark Carman, Ben Carterette, Charles
  L.~A. Clarke, Kevyn Collins-Thompson, Nick Craswell, W.~Bruce Croft, J.~Shane
  Culpepper, Jeff Dalton, Gianluca Demartini, Fernado Diaz, Laura Dietz, Susan
  Dumais, Carsten Eickhoff, Nicola Ferro, Norbert Fuhr, Shlomo Geva, Claudia
  Hauff, David Hawking, Hideo Joho, Gareth Jones, Jaap Kamps, Noriko Kando,
  Diane Kelly, Jaewon Kim, Julia Kiseleva, Yiqun Liu, Xiaolu Lu, Stefano
  Mizzaro, Alistair Moffat, Jian-Yun Nie, Alexandra Olteanu, Iadh Ounis, Filip
  Radlinski, Maarten de~Rijke, Mark Sanderson, Falk Scholer, Laurianne Sitbon,
  Mark Smucker, Ian Soboroff, Damiano Spina, Torsten Suel, James Thom, Paul
  Thomas, Andrew Trotman, Ellen Voorhees, Arjen~P. de~Vries, Emine Yilmaz, and
  Guido Zuccon.
\newblock Report from the third strategic workshop on information retrieval in
  {Lorne} ({SWIRL} 2018).
\newblock \emph{SIGIR Forum}, 52:\penalty0 34--90, June 2018.

\bibitem[Bahri et~al.(2024)Bahri, Dyer, Kaplan, Lee, and
  Sharma]{bahri-2024-explaining}
Yasaman Bahri, Ethan Dyer, Jared Kaplan, Jaehoon Lee, and Utkarsh Sharma.
\newblock Explaining neural scaling laws.
\newblock \emph{Proceedings of the National Academy of Sciences}, 121\penalty0
  (27):\penalty0 e2311878121, 2024.

\bibitem[Bai et~al.(2024)Bai, da~Silva~Torres, Fensel, Metze, and
  Dewulf]{bai-2024-inferring}
Nan Bai, Ricardo da~Silva~Torres, Anna Fensel, Tamara Metze, and Art Dewulf.
\newblock Inferring climate change stances from multimodal tweets.
\newblock In \emph{Proceedings of the 47th International ACM SIGIR Conference
  on Research and Development in Information Retrieval}, pages 2467--2471.
  Association for Computing Machinery, 2024.

\bibitem[Balog(2018)]{balog-2018-entity}
Krisztian Balog.
\newblock \emph{Entity-Oriented Search}, volume~10 of \emph{Synthesis Lectures
  on Information Concepts, Retrieval, and Services}.
\newblock Morgan \& Claypool Publishers, 2018.

\bibitem[Barnett and O'Neill(2013)]{barnett-2013-minimising}
Jon Barnett and Saffron~J. O'Neill.
\newblock Minimising the risk of maladaptation.
\newblock In \emph{Climate Adaptation Futures}, chapter~7, pages 87--93. John
  Wiley \& Sons, Ltd, 2013.

\bibitem[Berrang-Ford et~al.(2021)Berrang-Ford, Siders, Lesnikowski, Fischer,
  Callaghan, Haddaway, Mach, Araos, Shah, Wannewitz, Doshi, Leiter, Matavel,
  Musah-Surugu, Wong-Parodi, Antwi-Agyei, Ajibade, Chauhan, Kakenmaster, Grady,
  Chalastani, Jagannathan, Galappaththi, Sitati, Scarpa, Totin, Davis,
  Hamilton, Kirchhoff, Kumar, Pentz, Simpson, Theokritoff, Deryng, Reckien,
  Zavaleta-Cortijo, Ulibarri, Segnon, Khavhagali, Shang, Zvobgo, Zommers, Xu,
  Williams, Canosa, van Maanen, van Bavel, van Aalst, Turek-Hankins, Trivedi,
  Trisos, Thomas, Thakur, Templeman, Stringer, Sotnik, Sjostrom, Singh,
  Si{\~n}a, Shukla, Sardans, Salubi, Safaee~Chalkasra, Ruiz-D{\'\i}az,
  Richards, Pokharel, Petzold, Penuelas, Pelaez~Avila, Murillo, Ouni, Niemann,
  Nielsen, New, Nayna~Schwerdtle, Nagle~Alverio, Mullin, Mullenite, Mosurska,
  Morecroft, Minx, Maskell, Nunbogu, Magnan, Lwasa, Lukas-Sithole, Lissner,
  Lilford, Koller, Jurjonas, Joe, Huynh, Hill, Hernandez, Hegde, Hawxwell,
  Harper, Harden, Haasnoot, Gilmore, Gichuki, Gatt, Garschagen, Ford, Forbes,
  Farrell, Enquist, Elliott, Duncan, Coughlan~de Perez, Coggins, Chen,
  Campbell, Browne, Bowen, Biesbroek, Bhatt, Bezner~Kerr, Barr, Baker, Austin,
  Arotoma-Rojas, Anderson, Ajaz, Agrawal, and
  Abu]{berrang-ford-2021-systematic}
Lea Berrang-Ford, A.~R. Siders, Alexandra Lesnikowski, Alexandra~Paige Fischer,
  Max~W. Callaghan, Neal~R. Haddaway, Katharine~J. Mach, Malcolm Araos,
  Mohammad Aminur~Rahman Shah, Mia Wannewitz, Deepal Doshi, Timo Leiter,
  Custodio Matavel, Justice~Issah Musah-Surugu, Gabrielle Wong-Parodi, Philip
  Antwi-Agyei, Idowu Ajibade, Neha Chauhan, William Kakenmaster, Caitlin Grady,
  Vasiliki~I. Chalastani, Kripa Jagannathan, Eranga~K. Galappaththi, Asha
  Sitati, Giulia Scarpa, Edmond Totin, Katy Davis, Nikita~Charles Hamilton,
  Christine~J. Kirchhoff, Praveen Kumar, Brian Pentz, Nicholas~P. Simpson,
  Emily Theokritoff, Delphine Deryng, Diana Reckien, Carol Zavaleta-Cortijo,
  Nicola Ulibarri, Alcade~C. Segnon, Vhalinavho Khavhagali, Yuanyuan Shang,
  Luckson Zvobgo, Zinta Zommers, Jiren Xu, Portia~Adade Williams,
  Ivan~Villaverde Canosa, Nicole van Maanen, Bianca van Bavel, Maarten van
  Aalst, Lyn{\'e}e~L. Turek-Hankins, Hasti Trivedi, Christopher~H. Trisos,
  Adelle Thomas, Shinny Thakur, Sienna Templeman, Lindsay~C. Stringer, Garry
  Sotnik, Kathryn~Dana Sjostrom, Chandni Singh, Mariella~Z. Si{\~n}a, Roopam
  Shukla, Jordi Sardans, Eunice~A. Salubi, Lolita~Shaila Safaee~Chalkasra,
  Raquel Ruiz-D{\'\i}az, Carys Richards, Pratik Pokharel, Jan Petzold, Josep
  Penuelas, Julia Pelaez~Avila, Julia B.~Pazmino Murillo, Souha Ouni, Jennifer
  Niemann, Miriam Nielsen, Mark New, Patricia Nayna~Schwerdtle, Gabriela
  Nagle~Alverio, Cristina~A. Mullin, Joshua Mullenite, Anuszka Mosurska,
  Mike~D. Morecroft, Jan~C. Minx, Gina Maskell, Abraham~Marshall Nunbogu,
  Alexandre~K. Magnan, Shuaib Lwasa, Megan Lukas-Sithole, Tabea Lissner, Oliver
  Lilford, Steven~F. Koller, Matthew Jurjonas, Elphin~Tom Joe, Lam T.~M. Huynh,
  Avery Hill, Rebecca~R. Hernandez, Greeshma Hegde, Tom Hawxwell, Sherilee
  Harper, Alexandra Harden, Marjolijn Haasnoot, Elisabeth~A. Gilmore, Leah
  Gichuki, Alyssa Gatt, Matthias Garschagen, James~D. Ford, Andrew Forbes,
  Aidan~D. Farrell, Carolyn A.~F. Enquist, Susan Elliott, Emily Duncan, Erin
  Coughlan~de Perez, Shaugn Coggins, Tara Chen, Donovan Campbell, Katherine~E.
  Browne, Kathryn~J. Bowen, Robbert Biesbroek, Indra~D. Bhatt, Rachel
  Bezner~Kerr, Stephanie~L. Barr, Emily Baker, Stephanie~E. Austin, Ingrid
  Arotoma-Rojas, Christa Anderson, Warda Ajaz, Tanvi Agrawal, and
  Thelma~Zulfawu Abu.
\newblock A systematic global stocktake of evidence on human adaptation to
  climate change.
\newblock \emph{Nature Climate Change}, 11\penalty0 (11):\penalty0 989--1000,
  November 2021.

\bibitem[Callaghan et~al.(2021)Callaghan, Schleussner, Nath, Lejeune, Knutson,
  Reichstein, Hansen, Theokritoff, Andrijevic, Brecha, Hegarty, Jones, Lee,
  Lucas, van Maanen, Menke, Pfleiderer, Yesil, and
  Minx]{callaghan-2021-machine-learning-based}
Max Callaghan, Carl-Friedrich Schleussner, Shruti Nath, Quentin Lejeune,
  Thomas~R. Knutson, Markus Reichstein, Gerrit Hansen, Emily Theokritoff,
  Marina Andrijevic, Robert~J. Brecha, Michael Hegarty, Chelsea Jones, Kaylin
  Lee, Agathe Lucas, Nicole van Maanen, Inga Menke, Peter Pfleiderer, Burcu
  Yesil, and Jan~C. Minx.
\newblock Machine-learning-based evidence and attribution mapping of 100,000
  climate impact studies.
\newblock \emph{Nature Climate Change}, 11:\penalty0 966--972, 2021.

\bibitem[Callaghan et~al.(2020)Callaghan, Minx, and
  Forster]{callaghan-2020-topography}
Max~W. Callaghan, Jan~C. Minx, and Piers~M. Forster.
\newblock A topography of climate change research.
\newblock \emph{Nature Climate Change}, 10:\penalty0 118--123, 2020.

\bibitem[Conway et~al.(2020)Conway, Nicholls, Brown, Tebboth, Adger, Ahmad,
  Biemans, Crick, Lutz, Campos, Said, Singh, Zaroug, Ludi, New, and
  Wester]{conway-2019-need}
Declan Conway, Robert~J. Nicholls, Sally Brown, Mark G.~L. Tebboth,
  William~Neil Adger, Bashir Ahmad, Hester Biemans, Florence Crick, Arthur~F.
  Lutz, Ricardo Safra~De Campos, Mohammed Said, Chandni Singh, Modathir
  Abdalla~Hassan Zaroug, Eva Ludi, Mark New, and Philippus Wester.
\newblock The need for bottom-up assessments of climate risks and adaptation in
  climate-sensitive regions.
\newblock \emph{Nature Climate Change}, 9:\penalty0 503--511, 2020.

\bibitem[Cooper et~al.(2018)Cooper, Booth, Varley-Campbell, Britten, and
  Garside]{cooper-2018-defining}
Chris Cooper, Andrew Booth, Jo~Varley-Campbell, Nicky Britten, and Ruth
  Garside.
\newblock Defining the process to literature searching in systematic reviews: A
  literature review of guidance and supporting studies.
\newblock \emph{BMC Medical Research Methodology}, 18\penalty0 (1):\penalty0
  85, 2018.

\bibitem[De-Gol et~al.(2023)De-Gol, Le~Qu{\'e}r{\'e}, Smith, and Aubin
  Le~Qu{\'e}r{\'e}]{de-gol-2023-broadening}
Anthony~Jude De-Gol, Corinne Le~Qu{\'e}r{\'e}, Adam J.~P. Smith, and Marianne
  Aubin Le~Qu{\'e}r{\'e}.
\newblock Broadening scientific engagement and inclusivity in {IPCC} reports
  through collaborative technology platforms.
\newblock \emph{npj Climate Action}, 2\penalty0 (1):\penalty0 49, 2023.

\bibitem[Elliott et~al.(2017)Elliott, Synnot, Turner, Simmonds, Akl, McDonald,
  Salanti, Meerpohl, MacLehose, Hilton, Tovey, Shemilt, Thomas, Agoritsas,
  Hilton, Perron, Akl, Hodder, Pestridge, Albrecht, Horsley, Platt, Armstrong,
  Nguyen, Plovnick, Arno, Ivers, Quinn, Au, Johnston, Rada, Bagg, Jones,
  Ravaud, Boden, Kahale, Richter, Boisvert, Keshavarz, Ryan, Brandt,
  Kolakowsky-Hayner, Salama, Brazinova, Nagraj, Salanti, Buchbinder, Lasserson,
  Santaguida, Champion, Lawrence, Santesso, Chandler, Les, Sch{\"u}nemann,
  Charidimou, Leucht, Shemilt, Chou, Low, Sherifali, Churchill, Maas,
  Siemieniuk, Cnossen, MacLehose, Simmonds, Cossi, Macleod, Skoetz, Counotte,
  Marshall, Soares-Weiser, Craigie, Marshall, Srikanth, Dahm, Martin, Sullivan,
  Danilkewich, {Mart{\'\i}nez Garc{\'\i}a}, Synnot, Danko, Mavergames, Taylor,
  Donoghue, Maxwell, Thayer, Dressler, McAuley, Thomas, Egan, McDonald,
  Tritton, Elliott, McKenzie, Tsafnat, Elliott, Meerpohl, Tugwell, Etxeandia,
  Merner, Turgeon, Featherstone, Mondello, Turner, Foxlee, Morley, {van
  Valkenhoef}, Garner, Munafo, Vandvik, Gerrity, Munn, Wallace, Glasziou,
  Murano, Wallace, Green, Newman, Watts, Grimshaw, Nieuwlaat, Weeks, Gurusamy,
  Nikolakopoulou, Weigl, Haddaway, Noel-Storr, Wells, Hartling, O'Connor,
  Wiercioch, Hayden, Page, Wolfenden, Helfand, Pahwa, {Yepes Nu{\~n}ez},
  Higgins, Pardo, Yost, Hill, and Pearson]{elliott-2017-living}
Julian~H. Elliott, Anneliese Synnot, Tari Turner, Mark Simmonds, Elie~A. Akl,
  Steve McDonald, Georgia Salanti, Joerg Meerpohl, Harriet MacLehose, John
  Hilton, David Tovey, Ian Shemilt, James Thomas, Thomas Agoritsas, John
  Hilton, Caroline Perron, Elie Akl, Rebecca Hodder, Charlotte Pestridge,
  Lauren Albrecht, Tanya Horsley, Joanne Platt, Rebecca Armstrong, Phi~Hung
  Nguyen, Robert Plovnick, Anneliese Arno, Noah Ivers, Gail Quinn, Agnes Au,
  Renea Johnston, Gabriel Rada, Matthew Bagg, Arwel Jones, Philippe Ravaud,
  Catherine Boden, Lara Kahale, Bernt Richter, Isabelle Boisvert, Homa
  Keshavarz, Rebecca Ryan, Linn Brandt, Stephanie~A. Kolakowsky-Hayner, Dina
  Salama, Alexandra Brazinova, Sumanth~Kumbargere Nagraj, Georgia Salanti,
  Rachelle Buchbinder, Toby Lasserson, Lina Santaguida, Chris Champion, Rebecca
  Lawrence, Nancy Santesso, Jackie Chandler, Zbigniew Les, Holger~J.
  Sch{\"u}nemann, Andreas Charidimou, Stefan Leucht, Ian Shemilt, Roger Chou,
  Nicola Low, Diana Sherifali, Rachel Churchill, Andrew Maas, Reed Siemieniuk,
  Maryse~C. Cnossen, Harriet MacLehose, Mark Simmonds, Marie-Joelle Cossi,
  Malcolm Macleod, Nicole Skoetz, Michel Counotte, Iain Marshall, Karla
  Soares-Weiser, Samantha Craigie, Rachel Marshall, Velandai Srikanth, Philipp
  Dahm, Nicole Martin, Katrina Sullivan, Alanna Danilkewich, Laura
  {Mart{\'\i}nez Garc{\'\i}a}, Anneliese Synnot, Kristen Danko, Chris
  Mavergames, Mark Taylor, Emma Donoghue, Lara~J. Maxwell, Kris Thayer, Corinna
  Dressler, James McAuley, James Thomas, Cathy Egan, Steve McDonald, Roger
  Tritton, Julian Elliott, Joanne McKenzie, Guy Tsafnat, Sarah~A. Elliott,
  Joerg Meerpohl, Peter Tugwell, Itziar Etxeandia, Bronwen Merner, Alexis
  Turgeon, Robin Featherstone, Stefania Mondello, Tari Turner, Ruth Foxlee,
  Richard Morley, Gert {van Valkenhoef}, Paul Garner, Marcus Munafo, Per
  Vandvik, Martha Gerrity, Zachary Munn, Byron Wallace, Paul Glasziou, Melissa
  Murano, Sheila~A. Wallace, Sally Green, Kristine Newman, Chris Watts, Jeremy
  Grimshaw, Robby Nieuwlaat, Laura Weeks, Kurinchi Gurusamy, Adriani
  Nikolakopoulou, Aaron Weigl, Neal Haddaway, Anna Noel-Storr, George Wells,
  Lisa Hartling, Annette O'Connor, Wojtek Wiercioch, Jill Hayden, Matthew Page,
  Luke Wolfenden, Mark Helfand, Manisha Pahwa, Juan~Jos{\'e} {Yepes Nu{\~n}ez},
  Julian Higgins, Jordi~Pardo Pardo, Jennifer Yost, Sophie Hill, and Leslea
  Pearson.
\newblock Living systematic review: 1. introduction---the why, what, when, and
  how.
\newblock \emph{Journal of Clinical Epidemiology}, 91:\penalty0 23--30, 2017.

\bibitem[Frew et~al.(2024)Frew, Nelson, and Weigle]{frew-2024-retrogressive}
Lesley Frew, Michael~L. Nelson, and Michele~C. Weigle.
\newblock Retrogressive document manipulation of {US} federal environmental
  websites.
\newblock In \emph{Proceedings of the 33rd ACM International Conference on
  Information and Knowledge Management}, pages 3762--3766, 2024.

\bibitem[Ghosh et~al.(2022)Ghosh, Harihar, and Jain]{ghosh-2022-co-development}
Arunabha Ghosh, Nandini Harihar, and Prayank Jain.
\newblock Co-development of technologies of the future.
\newblock Technical report, Stockholm+50 background paper series. Stockholm
  Environment Institute, 2022.

\bibitem[Hausfather et~al.(2022)Hausfather, Marvel, Schmidt, Nielsen-Gammon,
  and Zelinka]{hausfather-2022-climate}
Zeke Hausfather, Kate Marvel, Gavin~A. Schmidt, John~W. Nielsen-Gammon, and
  Mark Zelinka.
\newblock Climate simulations: recognize the `hot model' problem.
\newblock \emph{Nature}, 605:\penalty0 26--29, 2022.

\bibitem[Hempel et~al.(2014)Hempel, Taylor, Marshall, Miake-Lye, Beroes,
  Shanman, Solloway, and Shekelle]{hempel-2014-evidence}
Susanne Hempel, Stephanie~L. Taylor, Nell~J. Marshall, Isomi~M. Miake-Lye,
  Jessica~M. Beroes, Roberta Shanman, Michele~R. Solloway, and Paul~G.
  Shekelle.
\newblock Evidence map of mindfulness.
\newblock Technical report, Department of Veterans Affairs, 2014.

\bibitem[Hocking et~al.(2023)Hocking, Parkinson, Adams, Molding~Nielsen, Ang,
  and de~Carvalho~Gomes]{hocking-2023-overcoming}
Lucy Hocking, Sarah Parkinson, Avery Adams, Emmanuel Molding~Nielsen, Cecilia
  Ang, and Helena de~Carvalho~Gomes.
\newblock Overcoming the challenges of using automated technologies for public
  health evidence synthesis.
\newblock \emph{BMC Public Health}, 28\penalty0 (45):\penalty0 2300183, 2023.

\bibitem[{IPCC}(2022{\natexlab{a}})]{ipcc-2022-climate}
{IPCC}.
\newblock \emph{Climate Change 2022: Impacts, Adaptation and Vulnerability}.
\newblock Cambridge University Press, 2022{\natexlab{a}}.
\newblock H.-O. P{\"o}rtner, D.C. Roberts, M. Tignor, E.S. Poloczanska, K.
  Mintenbeck, A. Alegr{\'\i}a, M. Craig, S. Langsdorf, S. L{\"o}schke, V.
  M{\"o}ller, A. Okem and B. Rama, editors.

\bibitem[{IPCC}(2022{\natexlab{b}})]{ipcc-2022-summary}
{IPCC}.
\newblock Summary for policymakers.
\newblock In H.-O. P{\"o}rtner, D.C. Roberts, E.S. Poloczanska, K.~Mintenbeck,
  M.~Tignor, A.~Alegr{\'\i}a, M.~Craig, S.~Langsdorf, S.~L{\"o}schke,
  V.~M{\"o}ller, and A.~Okem, editors, \emph{Climate Change 2022: Impacts,
  Adaptation and Vulnerability. Contribution of Working Group II to the Sixth
  Assessment Report of the Intergovernmental Panel on Climate Change}, pages
  3--33. Cambridge University Press, 2022{\natexlab{b}}.

\bibitem[IPCC, 2024()]{ipcc-2024-wgii}
IPCC, 2024.
\newblock Working group {II} impacts, adaptation and vulnerability.
\newblock \url{https://www.ipcc.ch/working-group/wg2/}, 2024.

\bibitem[Janowicz et~al.(2022)Janowicz, Hitzler, Li, Rehberger, Schildhauer,
  Zhu, Shimizu, Fisher, Cai, Mai, et~al.]{janowicz-2022-know}
Krzysztof Janowicz, Pascal Hitzler, Wenwen Li, Dean Rehberger, Mark
  Schildhauer, Rui Zhu, Cogan Shimizu, Colby Fisher, Ling Cai, Gengchen Mai,
  et~al.
\newblock {Know, Know Where, KnowWhereGraph}: A densely connected, cross-domain
  knowledge graph and geo-enrichment service stack for applications in
  environmental intelligence.
\newblock \emph{AI Magazine}, 43\penalty0 (1):\penalty0 30--39, 2022.

\bibitem[Joe et~al.(2024)Joe, Koneru, and Kirchhoff]{joe-2024-assessing}
Elphin~Tom Joe, Sai Koneru, and Christine~J. Kirchhoff.
\newblock Assessing the effectiveness of {GPT}-4o in climate change evidence
  synthesis and systematic assessments: Preliminary insights.
\newblock In \emph{ClimateNLP 2024: ACL 2024 Workshop on Natural Language
  Processing meets Climate Change}. ACL, August 2024.

\bibitem[Kanoulas et~al.(2019)Kanoulas, Li, Azzopardi, and
  Spijker]{kanoulas-2019-clef}
Evangelos Kanoulas, Dan Li, Leif Azzopardi, and Rene Spijker.
\newblock {CLEF} 2019 technology assisted reviews in empirical medicine
  overview.
\newblock In \emph{Working Notes of CLEF 2019 --- Conference and Labs of the
  Evaluation Forum}, volume 2380, page 250, 2019.
\newblock URL \url{https://ceur-ws.org/Vol-2380/paper_250.pdf}.

\bibitem[Kusa et~al.(2023)Kusa, Mendoza, Samwald, Knoth, and
  Hanbury]{kusa-2023-csmed}
Wojciech Kusa, Oscar~E. Mendoza, Matthias Samwald, Petr Knoth, and Allan
  Hanbury.
\newblock {CSMeD}: Bridging the dataset gap in automated citation screening for
  systematic literature reviews.
\newblock \emph{arXiv preprint arXiv:2311.12474}, 2023.

\bibitem[Lansbury et~al.(2023)Lansbury, Moggridge, Creamer, Ireland, Buckley,
  Evans, Milsom, Pecl, and Mosby]{lansbury-2023-aboriginal}
Nina Lansbury, Brad Moggridge, Sandra Creamer, Lillian Ireland, Lisa Buckley,
  Geoff Evans, Olivia Milsom, Gretta Pecl, and Vinnitta Mosby.
\newblock Aboriginal and {Torres Strait Islander Peoples'} voices and
  engagement in the {Intergovernmental Panel on Climate Change}: {Advice} to
  inform the {Australian Government} towards {IPCC Assessment Report 7}.
\newblock An independent report commissioned by the Australian Government
  (Department of Climate Change, Energy, the Environment and Water), Canberra.
  \url{https://public-health.uq.edu.au/files/25162/IPCC-Voices-Report.pdf},
  2023.

\bibitem[Lewis et~al.(2020)Lewis, Abdilla, Arista, Baker, Benesiinaabandan,
  Brown, Cheung, Coleman, Cordes, Davison, Duncan, Garzon, Harrell, Jones,
  Kealiikanakaoleohaililani, Kelleher, Kite, Lagon, Leigh, Levesque, Mahelona,
  Moses, Nahuewai, Noe, Olson, Parker~Jones, Running~Wolf, Running~Wolf, Silva,
  Fragnito, and Whaanga]{lewis-2020-indigenous}
Jason~Edward Lewis, Angie Abdilla, Noelani Arista, Kaipulaumakaniolono Baker,
  Scott Benesiinaabandan, Michelle Brown, Melanie Cheung, Meredith Coleman,
  Ashley Cordes, Joel Davison, K{\=u}pono Duncan, Sergio Garzon, D.~Fox
  Harrell, Peter-Lucas Jones, Kekuhi Kealiikanakaoleohaililani, Megan Kelleher,
  Suzanne Kite, Olin Lagon, Jason Leigh, Maroussia Levesque, Keoni Mahelona,
  Caleb Moses, Isaac~('Ika'aka) Nahuewai, Kari Noe, Danielle Olson, '{\=O}iwi
  Parker~Jones, Caroline Running~Wolf, Michael Running~Wolf, Marlee Silva,
  Skawennati Fragnito, and H{\=e}mi Whaanga.
\newblock Indigenous protocol and artificial intelligence position paper.
\newblock Technical report, Aboriginal Territories in Cyberspace, Honolulu, HI,
  2020.
\newblock Edited by Jason Edward Lewis. English Language Version of ``Ka`ina
  Hana `{\=O}iwi a me ka Waihona `Ike Hakuhia Pepa K{\=u}lana'' available at:
  \url{https://spectrum.library.concordia.ca/id/eprint/990094/}.

\bibitem[Lyu et~al.(2025)Lyu, Yan, Wang, Yin, Ren, de~Rijke, and
  Ren]{lyu-2025-macpo}
Yougang Lyu, Lingyong Yan, Zihan Wang, Dawei Yin, Pengjie Ren, Maarten
  de~Rijke, and Zhaochun Ren.
\newblock {MACPO}: Weak-to-strong alignment via multi-agent contrastive
  preference optimization.
\newblock In \emph{The Thirteenth International Conference on Learning
  Representations}, April 2025.

\bibitem[Mallick et~al.(2024)Mallick, Murphy, Bergerson, Verner, Hutchison, and
  Levy]{mallick-2024-analyzing}
Tanwi Mallick, John Murphy, Joshua~David Bergerson, Duane~R. Verner, John~K.
  Hutchison, and Leslie-Anne Levy.
\newblock Analyzing regional impacts of climate change using natural language
  processing techniques.
\newblock \emph{arXiv preprint arXiv:2401.06817}, 2024.

\bibitem[Mastrandrea et~al.(2010)Mastrandrea, Field, Stocker, Edenhofer, Ebi,
  Frame, Held, Kriegler, Mach, Matschoss, Plattner, Yohe, and
  Zwiers]{mastrandrea-2010-guidance}
Michael~D. Mastrandrea, Christopher~B. Field, Thomas~F. Stocker, Ottmar
  Edenhofer, Kristie~L. Ebi, David~J. Frame, Hermann Held, Elmar Kriegler,
  Katharine~J. Mach, Patrick~R. Matschoss, Gian-Kasper Plattner, Gary~W. Yohe,
  and Francis~W. Zwiers.
\newblock Guidance note for lead authors of the {IPCC} fifth assessment report
  on consistent treatment of uncertainties.
\newblock Technical report, Intergovernmental Panel on Climate Change (IPCC),
  2010.
\newblock
  \url{https://www.ipcc.ch/site/assets/uploads/2017/08/AR5_Uncertainty_Guidance_Note.pdf}.

\bibitem[Meij et~al.(2013)Meij, Balog, and Odijk]{meij-2013-entity}
Edgar Meij, Krisztian Balog, and Daan Odijk.
\newblock Entity linking and retrieval.
\newblock In \emph{Proceedings of the 36th International ACM SIGIR Conference
  on Research and Development in Information Retrieval}, page 1127, 2013.

\bibitem[Minx et~al.(2017)Minx, Callaghan, Lamb, Garard, and
  Edenhofer]{minx-2017-learning}
Jan~C. Minx, Max Callaghan, William~F. Lamb, Jennifer Garard, and Ottmar
  Edenhofer.
\newblock Learning about climate change solutions in the {IPCC} and beyond.
\newblock \emph{Environmental Science \& Policy}, 77:\penalty0 252--259, 2017.

\bibitem[O'Neill et~al.(2022)O'Neill, van Aalst, Zaiton~Ibrahim, Berrang-Ford,
  Bhadwal, Buhaug, Diaz, Frieler, Garschagen, Magnan, Midgley, Mirzabaev,
  Thomas, and Warren]{oneill-2022-key}
Brian~C. O'Neill, Maarten van Aalst, Zelina Zaiton~Ibrahim, Lea Berrang-Ford,
  Suruchi Bhadwal, Halvard Buhaug, Delane Diaz, Katja Frieler, Matthias
  Garschagen, Alexandre~K. Magnan, Guy Midgley, Alisher Mirzabaev, Adelle
  Thomas, and Rachel Warren.
\newblock Key risks across sectors and regions.
\newblock In H.-O. P{\"o}rtner, D.C. Roberts, M.~Tignor, E.S. Poloczanska,
  K.~Mintenbeck, A.~Alegr{\'\i}a, M.~Craig, S.~Langsdorf, S.~L{\"o}schke,
  V.~M{\"o}ller, A.~Okem, and B.~Rama, editors, \emph{Climate Change 2022:
  Impacts, Adaptation and Vulnerability. Contribution of Working Group II to
  the Sixth Assessment Report of the Intergovernmental Panel on Climate
  Change}, pages 2411--2538. Cambridge University Press, 2022.

\bibitem[P{\"o}rtner et~al.(2022)P{\"o}rtner, Roberts, Adams, Adelekan, Adler,
  Adrian, Aldunce, Ali, Begum, Bednar-Friedl, Kerr, Biesbroek, Birkmann, Bowen,
  Caretta, Carnicer, Castellanos, Cheong, Chow, Ciss{\'e}, Clayton, Constable,
  Cooley, Costello, Craig, Cramer, Dawson, Dodman, Efitre, Garschagen, Gilmore,
  Glavovic, Gutzler, Haasnoot, Harper, Hasegawa, Hayward, Hicke, Hirabayashi,
  Huang, Kalaba, Kiessling, Kitoh, Lasco, Lawrence, Lemos, Lempert, Lennard,
  Ley, Lissner, Liu, Liwenga, Lluch-Cota, Loeschke, Lucatello, Luo, Mintenbeck,
  Mirzabaev, Moeller, Vale, Morecroft, Mortsch, Mukherji, Mustonen, Mycoo,
  Nalau, New, Okem, Ometto, O'Neill, Pandey, Parmesan, Pelling, Pinho,
  Pinnegar, Poloczanska, Prakash, Preston, Racault, Reckien, Revi, Rose,
  Schipper, Schmidt, Schoeman, Shaw, Simpson, Singh, Solecki, Stringer, Totin,
  Trisos, Trisurat, van Aalst, Viner, Wairiu, Warren, Wester, Wrathall, and
  Ibrahim]{portner-2022-technical}
Hans P{\"o}rtner, Debra~C. Roberts, Helen Adams, Ibidun Adelekan, Carolina
  Adler, Rita Adrian, Paulina Aldunce, Elham Ali, Rawshan~Ara Begum, Birgit
  Bednar-Friedl, Rachel~Bezner Kerr, Robbert Biesbroek, Joern Birkmann, Kathryn
  Bowen, Martina~Angela Caretta, Jofre Carnicer, Edwin Castellanos, Tae~Sung
  Cheong, Winston Chow, Gu{\'e}ladio Ciss{\'e}, Susan Clayton, Andrew
  Constable, Sarah~R. Cooley, Mark~John Costello, Marlies Craig, Wolfgang
  Cramer, Richard Dawson, David Dodman, Jackson Efitre, Matthias Garschagen,
  Elisabeth Gilmore, Bruce Glavovic, David Gutzler, Marjolijn Haasnoot,
  Sherilee Harper, Toshihiro Hasegawa, Bronwyn Hayward, Jeffrey Hicke, Yukiko
  Hirabayashi, Cunrui Huang, Kanungwe Kalaba, Wolfgang Kiessling, Akio Kitoh,
  Rodel Lasco, Judy Lawrence, Maria~Fernanda Lemos, Robert Lempert, Christopher
  Lennard, Debora Ley, Tabea Lissner, Qiyong Liu, Emma Liwenga, Salvador
  Lluch-Cota, Sina Loeschke, Simone Lucatello, Yong Luo, Brendan Mackey~Katja
  Mintenbeck, Alisher Mirzabaev, Vincent Moeller, Mariana~Moncassim Vale, Mike
  Morecroft, Linda Mortsch, Aditi Mukherji, Tero Mustonen, Michelle Mycoo,
  Johanna Nalau, Mark New, Andrew Okem, Jean~Pierre Ometto, Brian O'Neill,
  Rajiv Pandey, Camille Parmesan, Mark Pelling, Patricia~Fernanda Pinho, John
  Pinnegar, Elvira~S. Poloczanska, Anjal Prakash, Benjamin Preston, Marie-Fanny
  Racault, Diana Reckien, Aromar Revi, Steven~K. Rose, E.~Lisa~F. Schipper,
  Daniela Schmidt, David Schoeman, Rajib Shaw, Nicholas~P. Simpson, Chandni
  Singh, William Solecki, Lindsay Stringer, Edmond Totin, Christopher Trisos,
  Yongyut Trisurat, Maarten van Aalst, David Viner, Morgan Wairiu, Rachel
  Warren, Philippus Wester, David Wrathall, and Zelina~Zaiton Ibrahim.
\newblock Technical summary.
\newblock In H.-O. P{\"o}rtner, D.C. Roberts, E.S. Poloczanska, K.~Mintenbeck,
  M.~Tignor, A.~Alegr{\'\i}a, M.~Craig, S.~Langsdorf, S.~L{\"o}schke,
  V.~M{\"o}ller, and A.~Okem, editors, \emph{Climate Change 2022: Impacts,
  Adaptation and Vulnerability. Contribution of Working Group II to the Sixth
  Assessment Report of the Intergovernmental Panel on Climate Change}, pages
  37--118. Cambridge University Press, 2022.

\bibitem[Rajapakse et~al.(2024)Rajapakse, Yates, and
  de~Rijke]{rajapakse-2024-simple}
Thilina Rajapakse, Andrew Yates, and Maarten de~Rijke.
\newblock Simple transformers: Open-source for all.
\newblock In \emph{SIGIR-AP 2024: Proceedings of the 2024 Annual International
  ACM SIGIR Conference on Research and Development in Information Retrieval in
  the Asia Pacific Region}, pages 209--215. ACM, December 2024.

\bibitem[Roberts et~al.(2021)Roberts, Carlson, O'Sullivan, Day, Rey, Kennedy,
  Bakic, and Farrell]{roberts-2021-guide}
Zac Roberts, Bronwyn Carlson, Sandy O'Sullivan, Madi Day, Jo~Rey, Tristan
  Kennedy, Tetei Bakic, and Andrew Farrell.
\newblock A guide to writing and speaking about {Indigenous People} in
  {Australia}.
\newblock Technical report, Macquarie University, 2021.
\newblock \url{https://doi.org/10.25949/5tfk-5113}.

\bibitem[Scells(2021)]{scells-2021-query}
Harrisen Scells.
\newblock \emph{Query Automation for Systematic Reviews}.
\newblock PhD thesis, School of Information Technology and Electrical
  Engineering, The University of Queensland, 2021.

\bibitem[Schleussner and Fyson(2020)]{schleusssner-2020-scenarios}
Carl-Friedrich Schleussner and Claire~L. Fyson.
\newblock Scenarios science needed in {UNFCCC} periodic review.
\newblock \emph{Nature Climate Change}, 10:\penalty0 272, 2020.

\bibitem[Shaib et~al.(2023)Shaib, Li, Joseph, Marshall, Li, and
  Wallace]{shaib-2023-summarizing}
Chantal Shaib, Millicent Li, Sebastian Joseph, Iain Marshall, Junyi~Jessy Li,
  and Byron Wallace.
\newblock Summarizing, simplifying, and synthesizing medical evidence using
  {GPT}-3 (with varying success).
\newblock In \emph{Proceedings of the 61st Annual Meeting of the Association
  for Computational Linguistics (Volume 2: Short Papers)}, pages 1387--1407.
  ACL, July 2023.

\bibitem[Sietsma et~al.(2021)Sietsma, Ford, Callaghan, and
  Minx]{sietsma-2021-progress}
Anne~J. Sietsma, James~D. Ford, Max~W. Callaghan, and Jan~C. Minx.
\newblock Progress in climate change adaptation research.
\newblock \emph{Environmental Research Letters}, 16\penalty0 (5):\penalty0
  054038, April 2021.

\bibitem[Sietsma et~al.(2024)Sietsma, Ford, and Minx]{sietsma-2024-next}
Anne~J. Sietsma, James~D. Ford, and Jan~C. Minx.
\newblock The next generation of machine learning for tracking adaptation
  texts.
\newblock \emph{Nature Climate Change}, 14:\penalty0 31--39, 2024.

\bibitem[Simmonds et~al.(2017)Simmonds, Salanti, McKenzie, Elliott, and behalf
  of~the Living Systematic Review~Network]{simmonds-2017-living}
Mark Simmonds, Georgia Salanti, Joanne McKenzie, Julian Elliott, and On~behalf
  of~the Living Systematic Review~Network.
\newblock Living systematic reviews: 3. {Statistical} methods for updating
  meta-analyses.
\newblock \emph{Journal of Clinical Epidemiology}, 91:\penalty0 38--46, 2017.

\bibitem[Stevenson and Bin-Hezam(2023)]{stevenson-2023-stopping}
Mark Stevenson and Reem Bin-Hezam.
\newblock Stopping methods for technology-assisted reviews based on point
  processes.
\newblock \emph{ACM Trans. Inf. Syst.}, 42\penalty0 (3), December 2023.

\bibitem[Tang et~al.(2024)Tang, Zhang, Ren, Guo, and
  de~Rijke]{tang-2024-recent-sigir}
Yubao Tang, Ruqing Zhang, Zhaochun Ren, Jiafeng Guo, and Maarten de~Rijke.
\newblock Recent advances in generative information retrieval.
\newblock In \emph{SIGIR 2024: 47th international ACM SIGIR Conference on
  Research and Development in Information Retrieval}, pages 3005--3008. ACM,
  July 2024.

\bibitem[Thakur et~al.(2021)Thakur, Reimers, R{\"u}ckl{\'e}, Srivastava, and
  Gurevych]{thakur-2021-beir}
Nandan Thakur, Nils Reimers, Andreas R{\"u}ckl{\'e}, Abhishek Srivastava, and
  Iryna Gurevych.
\newblock {BEIR}: A heterogeneous benchmark for zero-shot evaluation of
  information retrieval models.
\newblock In \emph{Thirty-fifth Conference on Neural Information Processing
  Systems Datasets and Benchmarks Track (Round 2)}, 2021.

\bibitem[Thomas et~al.(2017)Thomas, Noel-Storr, Marshall, Wallace, McDonald,
  Mavergames, Glasziou, Shemilt, Synnot, Turner, Elliott, and on~behalf of~the
  Living Systematic Review~Network]{thomas-2017-living}
James Thomas, Anna Noel-Storr, Iain Marshall, Byron Wallace, Steven McDonald,
  Chris Mavergames, Paul Glasziou, Ian Shemilt, Anneliese Synnot, Tari Turner,
  Julian Elliott, and on~behalf of~the Living Systematic Review~Network.
\newblock Living systematic reviews: 2. {Combining} human and machine effort.
\newblock \emph{Journal of Clinical Epidemiology}, 91:\penalty0 31--37, 2017.

\bibitem[Thomas et~al.(2024)Thomas, Spielman, Craswell, and
  Mitra]{thomas-2024-large}
Paul Thomas, Seth Spielman, Nick Craswell, and Bhaskar Mitra.
\newblock Large language models can accurately predict searcher preferences.
\newblock In \emph{SIGIR 2024: Proceedings of the 47th International ACM SIGIR
  Conference on Research and Development in Information Retrieval}, pages
  1930--1940. ACM, July 2024.

\bibitem[Thulke et~al.(2024)Thulke, Gao, Pelser, Brune, Jalota, Fok, Ramos, van
  Wyk, Nasir, Goldstein, Tragemann, Nguyen, Fowler, Stanco, Gabriel, Taylor,
  Moro, Tsymbalov, de~Waal, Matusov, Yaghi, Shihadah, Ney, Dugast, Dotan, and
  Erasmus]{thulke-2024-climategpt}
David Thulke, Yingbo Gao, Petrus Pelser, Rein Brune, Rricha Jalota, Floris Fok,
  Michael Ramos, Ian van Wyk, Abdallah Nasir, Hayden Goldstein, Taylor
  Tragemann, Katie Nguyen, Ariana Fowler, Andrew Stanco, Jon Gabriel, Jordan
  Taylor, Dean Moro, Evgenii Tsymbalov, Juliette de~Waal, Evgeny Matusov, Mudar
  Yaghi, Mohammad Shihadah, Hermann Ney, Christian Dugast, Jonathan Dotan, and
  Daniel Erasmus.
\newblock {ClimateGPT}: Towards {AI} synthesizing interdisciplinary research on
  climate change.
\newblock \emph{arXiv preprint arXiv:2401.09646}, January 2024.

\bibitem[Tian et~al.(2025)Tian, Li, Hu, Chen, Brook, Brubaker, Zhang, and
  Liljedahl]{tian-2025-advancing}
Yuanyuan Tian, Wenwen Li, Lei Hu, Xiao Chen, Michael Brook, Michael Brubaker,
  Fan Zhang, and Anna~K Liljedahl.
\newblock Advancing large language models for spatiotemporal and semantic
  association mining of similar environmental events.
\newblock \emph{Transactions in GIS}, 29\penalty0 (1):\penalty0 e13282, 2025.

\bibitem[{United Nations}(2024)]{ipcc-2024-about}
{United Nations}.
\newblock About the {IPCC}.
\newblock \url{https://www.ipcc.ch/about/}, 2024.

\bibitem[van~den Hurk et~al.(2024)van~den Hurk, de~Rijke, and
  Salim]{vandenhurk-2024-manila24}
Bart van~den Hurk, Maarten de~Rijke, and Flora~D. Salim.
\newblock {MANILA24}: {SIGIR 2024} workshop on information retrieval for
  climate impact.
\newblock In \emph{SIGIR 2024: 47th international ACM SIGIR Conference on
  Research and Development in Information Retrieval}, pages 3044--3046. ACM,
  July 2024.

\bibitem[Wang et~al.(2023)Wang, Scells, Koopman, and Zuccon]{wang-2023-can}
Shuai Wang, Harrisen Scells, Bevan Koopman, and Guido Zuccon.
\newblock Can {ChatGPT} write a good boolean query for systematic review
  literature search?
\newblock In \emph{Proceedings of the 46th International ACM SIGIR Conference
  on Research and Development in Information Retrieval}, pages 1426--1436.
  Association for Computing Machinery, July 2023.

\bibitem[Wang et~al.(2024)Wang, Scells, Zhuang, Potthast, Koopman, and
  Zuccon]{wang-2024-zero-shot}
Shuai Wang, Harrisen Scells, Shengyao Zhuang, Martin Potthast, Bevan Koopman,
  and Guido Zuccon.
\newblock Zero-shot generative large language models for systematic review
  screening automation.
\newblock In \emph{Advances in Information Retrieval}, pages 403--420.
  Springer, April 2024.

\bibitem[Wang et~al.(2025)Wang, Zhao, Lyu, Chen, de~Rijke, and
  Ren]{wang-2025-cooperative}
Zihan Wang, Ziqi Zhao, Yougang Lyu, Zhumin Chen, Maarten de~Rijke, and Zhaochun
  Ren.
\newblock A cooperative multi-agent framework for zero-shot named entity
  recognition.
\newblock In \emph{The Web Conference 2025}. ACM, April 2025.

\bibitem[Webersinke et~al.(2021)Webersinke, Kraus, Bingler, and
  Leippold]{webersinke-2021-climatebert}
Nicolas Webersinke, Mathias Kraus, Julia~Anna Bingler, and Markus Leippold.
\newblock {ClimateBert}: A pretrained language model for climate-related text.
\newblock \emph{arXiv preprint arXiv:2110.12010}, October 2021.

\bibitem[Xie et~al.(2024)Xie, Jiang, Mallick, Bergerson, Hutchison, Verner,
  Branham, Alexander, Ross, Feng, Levy, Su, and Taylor]{xie-2024-wildfiregpt}
Yangxinyu Xie, Bowen Jiang, Tanwi Mallick, Joshua~David Bergerson, John~K.
  Hutchison, Duane~R. Verner, Jordan Branham, M.~Ross Alexander, Robert~B.
  Ross, Yan Feng, Leslie-Anne Levy, Weijie Su, and Camillo~J. Taylor.
\newblock {WildfireGPT}: Tailored large language model for wildfire analysis.
\newblock \emph{arXiv preprint arXiv:2402.07877}, February 2024.

\bibitem[Zhang et~al.(2021)Zhang, Ren, and de~Rijke]{zhang-2021-human-machine}
Yangjun Zhang, Pengjie Ren, and Maarten de~Rijke.
\newblock A human-machine collaborative framework for evaluating malevolence in
  dialogues.
\newblock In \emph{ACL-IJCNLP 2021: The Joint Conference of the 59th Annual
  Meeting of the Association for Computational Linguistics and the 11th
  International Joint Conference on Natural Language Processing}, August 2021.

\end{thebibliography}

\end{document}